\documentclass[twocolumn,showpacs,amsmath,amssymb,aps,prl]{revtex4-1}
\usepackage{graphicx}
\usepackage{amsmath}
\newcommand{\beq}{\begin{equation}}
\newcommand{\eeq}{\end{equation}}
\newcommand{\bea}{\begin{eqnarray}}
\newcommand{\eea}{\end{eqnarray}}

\begin{document}

\title{Self-Consistent Field Theory of Disordered Block-Copolymers}

\author{Gabriele Migliorini}
\affiliation{Department of Mathematics and Statistics\\
University of Reading\\ 
RG6 6AX Reading, United Kingdom\\ }
\date{\today}

\begin{abstract}
We derive the mean-field theory of disordered block-copolymers 
composed of two monomeric species, combining Edwards' functional 
method with the replica technique of disordered systems.  
In the absence of disorder we recover the canonical self-consistent 
field theory of inhomogeneous polymers. In the presence of sequence 
disorder the theory can be regarded as a comprehensive, novel 
self-consistent treatment of copolymer melts, unifying the weak- 
and strong-inhomogeneous regimes. In particular, we study the stability 
of the microphase separation transition in a melt of diblock-copolymers, 
composed of $N$ monomers and  equal $A$- $B$-volume fraction, against 
the disruptive effect of disorder. 
We obtain a phase diagram in terms of the relevant parameters, namely 
the rescaled Flory-Huggins parameter $\chi N$ and disorder strength $p$. 
\end{abstract}
\maketitle

The interplay between topological and energetic effects in linear 
heteropolymeric systems is a central problem in soft-condensed matter 
and biophysics.
Heteropolymers composed of two monomeric species include for example 
block-copolymers, alternating $(AB)_n$ multiblock copolymers and  
random copolymers \cite{Helfand}\cite{deGennes}\cite{Obukhov}.
Polymer chains composed of several monomeric species include  
random heteropolymers, biopolymers as well as proteins, 
that are comprised of twenty amino-acids 
\cite{Orland}\cite{Shakhnovich}\cite{Grosberg}.

Block-copolymer melts undergo an order-disorder transition at which 
spatially periodic, coexisting microdomains of similar chemical 
composition form. In particular, mean-field theory 
\cite{Edwards}\cite{Leibler} predicts that a melt of diblock-copolymers 
with an equal $A$- and $B$-volume fraction undergoes a second order 
phase transition at which a lamellar phase (L) occurs. 
For different values of the $A$- and $B$-volume fraction, diblock-copolymers 
are known to present in addition other morphologies, characterized 
by periodic two- and three-dimensional spacial orderings, e.g. 
body-centered-cubic (bcc), hexagonally ordered cylindrical (hex) 
and bicontinous gyroid (G) \cite{Bates} and mean-field predicts all 
transitions to be first order in this case \cite{Schick}.
Different microphase orderings, induced by the appropriate choice of 
molecular weight and chain architecture during synthesis, have 
different thermal and mechanical properties. Consequently, 
block-copolymers find several applications in surfactant systems, 
as well as microelectronic systems and nanotechnology.

The phenomenon of microphase separation in systems of block-copolymer 
melts have been explained, within the weak-segregation 
approximation, more than thirty years ago \cite{Leibler}. 
A self-consistent, mean-field theory of block-copolymer systems for 
the gaussian chain model, beyond the weak-segregation approximation, 
has been formulated \cite{Helfand}\cite{Schick} and, despite the high 
computational cost required to solve the corresponding equations, 
it represents to date the leading method to investigate quantitatively
block-copolymer morphologies.
The phase diagram in terms of the Flory-Huggins parameter $ \chi$, 
the degree of polymerization $N$ and $A$-block volume fraction $f$ has 
been obtained and represents one of the crowning achievements of 
self-consistent field theory \cite{Schick}. 
\begin{figure}[ht]
\begin{center}
\includegraphics[width=0.3\textwidth,angle=270]{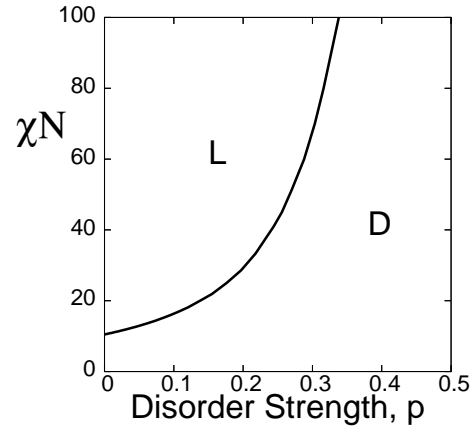}
\end{center}
\caption{ The phase diagram of a disordered diblock-copolymer melt 
with equal $A$- $B$-volume fraction $f=1/2$,  and disorder strength $p$, 
in terms of the relevant parameters, 
namely the rescaled Flory-Huggins parameter $ \chi N$ and disorder 
strength, $p$. The solid line represents the second-order phase 
boundary between the lamellar (L) and disordered (D) phases.}
\label{fig:fig1}
\end{figure}

In this letter we study the effect of quenched disorder 
on the equilibrium properties of block-copolymers \cite{Kuchanov}.
The equilibrium mean-field theory here introduced does not  
describe the effects of $self\mbox{-}generated$ 
disorder, that plays an important role in $dynamical$ mean-field 
theories of polymer melts and glassy systems \cite{Wolynes2}\cite{Zhang}.

The study of condensed-matter systems that possess quenched disorder  
\cite{Edwards2} represents a relatively old, yet rapidly evolving frontier 
of modern science. The mean-field theory of spin-glasses 
\cite{Sherrington}\cite{Parisi} has been extended to a variety of 
different problems, including the one of chemically disordered polymeric 
systems, firstly at the phenomenological level \cite{Wolynes}, and later 
from the microscopic point of view \cite{Shakhnovich2}. 
Quenched disorder in polymeric systems may occur to be fixed 
in space, when studying the behavior of macromolecules \cite{Vilgis}, 
manifolds \cite{Mezard} and directed polymers in random 
media \cite{Dotsenko} or might instead characterize the sequence 
of monomeric units.

The equilibrium properties of random $AB$ copolymer melts have been 
studied in the weak-segregation approximation, by means of the 
replica \cite{Fredrickson3} and cumulant expansion methods \cite{Fredrickson4}. 
In these two studies, the effect of correlated disorder and 
the presence of finite size $A$,$B$ multiblocks have been considered in detail. 
An equivalent expression for the Landau-type density functional free-energy 
has been obtained with both methods and the corresponding phase diagram, 
in the weak-segregation limit, has been discussed.
Related results have been obtained earlier, within the ground-state dominance 
approximation \cite{Shakhnovich2}, in the case of uncorrelated sequence 
disorder.

We derive the self-consistent theory of disordered heteropolymers, 
that includes both the weak- and strong-segregation regimes 
\cite{Fredrickson4}\cite{Semenov}, and apply it to the particular  
problem of disordered $AB$ diblock-copolymers. 

We consider the canonical formulation \cite{Helfand} for a melt of $n_p$ 
block-copolymer monodisperse chains composed of $N$ segments, 
where the index $i=1,\cdots,n_p$ labels each chain, 
where chains are characterized by $A$-block volume fraction $f$, 
where the total density of the system is $ \rho_o$, where 
$\{{ \bf r}_i(s) \}$ are the monomer cartesian positions 
and where $ \theta_i(s)= \pm 1$ are the occupation variables 
for segments $A$ and $B$ respectively, that occur to be 
disrupted with probability $p_{A,B}$, in each block independently.
The total density for segments of both type $A$ and $B$ and 
the relative density of the two species are defined as 
\bea
\hat{ \phi}({ \bf r})&=& \frac{N}{ 2 \rho_0} \sum_{i=1}^{n_p} \int_0^1 ds ~ \delta \big ( { \bf r}-{ \bf r}_i(s) \big ) \nonumber \\
\hat{m}({ \bf r}) &=& \frac{N}{ 2 \rho_0} \sum_{i=1}^{n_p} \int_0^1 ds ~ \theta_i(s) \delta \big ( { \bf r}-{ \bf r}_i(s) \big ) , 
\label{Densities}
\eea
and depend on the quenched random occupation variables $ \{ \theta_i \}$, 
where each specific realization encode the presence of uncorrelated 
sequence disorder within each block \cite{deGennes}\cite{Obukhov}. 
The model here introduced relates, without loss of continuity, 
the block-copolymer problem in the absence of disorder ($p_{A,B}=0$), 
with the problem of random $AB$ copolymers ($p_{A,B}=1/2$). 
The hamiltonian of the system is given by
\bea
&&{ \cal H}( \{ \theta \},\{ {\bf r} \}) = \frac{3}{2a^2N} \sum_{i=1}^{n_p} \int_0^1 ds|{ \bf r}_i(s)|^2 \nonumber \\
&&~~~~~+ \chi \rho_o \int d{ \bf r} [ \hat{m}^2({ \bf r})-\hat{ \phi}^2({ \bf r}) ], 
\label{Hamiltonian}
\eea
where $ \chi$ is the Flory-Huggins parameter, measuring the relative 
strength of the interaction between similar and dissimilar monomers, 
where we work in units of $k_BT$, and where $aN^{1/2}$ is the statistical 
segment length of the Gaussian chain model. 

We considered the case of uncorrelated disorder, so that the random 
occupation variables obey a binomial distribution in each block 
independently, with variance 
$ \langle \theta_i(s) \theta_j(s') \rangle = 4 \delta_{ij} \delta(s-s')p(1-p)$, where $p=p_{A,B}$ for $s<f$ and $s \ge f$ 
respectively, and where we assume that monomers $A$ and $B$ are 
represented by statistical segments with the same size $aN^{1/2}$. 
The order parameter $M$ is defined as the amplitude of the relative 
density of the two monomeric species, $ \hat{m}( { \bf r})$.
According to the Markov model of random copolymerization 
\cite{Fredrickson4}, each block is characterized 
by its length and disorder strength, chemically wired during 
synthesis, i.e. an average copolymer composition $p_{A,B}$ 
(or mole fraction of monomers of type $A$ in block $B$ and viceversa). 

\begin{figure}[ht]
\begin{center}
\includegraphics[width=0.45 \textwidth,angle=270]{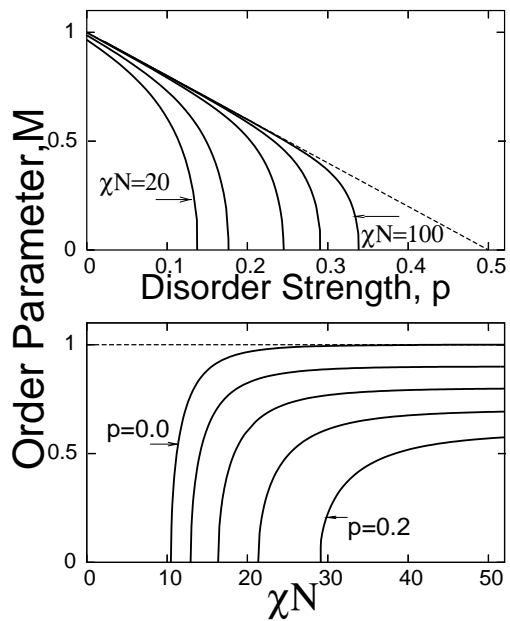}
\end{center}
\caption{The order parameter $M$ obtained from self-consistent 
field theory, for different values of the rescaled 
Flory-Huggins parameter $ \chi N$ and disorder strength. 
In the first plot (top) we show the order parameter $M$ 
as a function of disorder strength, for increasing values of the 
rescaled Flory-Huggins parameter $ \chi N = 20,25,40,80,100$ 
respectively.
In the second plot (bottom) we show the order parameter $M$ as a 
function of $ \chi N$, for increasing values of disorder strength 
$p=0,0.05,0.1,0.15,0.20$.
}
\label{fig:fig2}
\end{figure}

The two microscopic density operators in equation (\ref{Densities}) 
can be eliminated from the partition function, 
introducing the four, conjugate fields $\phi({ \bf r})$, $m({ \bf r})$, 
$w({ \bf r})$ and $v({ \bf r})$.
The incompressibility condition of the melt is enforced introducing the 
total chemical potential $ \xi({ \bf r})$. 
We average the partition function over the uncorrelated distribution of 
quenched random occupation variables in each block independently 
$ \langle \theta_i(s) \rangle =\pm( 1-2p_{A,B})$, according to the 
self-averaging hypothesis.
Differently from the problem of self-interacting heteropolymers 
composed of many monomeric species \cite{Shakhnovich}, where a 
replica symmetry broken scheme has to be introduced \cite{Grosberg}, 
the eigenvalue analysis of the $AB$ copolymer free-energy shows that 
the replica symmetric solution is stable \cite{Shakhnovich2}, so that 
the analytic continuation to the $n=0$ limit is a simple task 
\cite{Fredrickson4}. 
The self-consistent saddle point equations are given by 
$ w(r)= 2 \chi N \phi( { \bf r})+ \xi( { \bf r})$ and 
$v( { \bf r})= - 2 \chi N m({ \bf r})$. The conjugated density 
fields are given by
\bea
m({ \bf r})&=&\frac{1}{ { 2 \cal Q}} \int_0^1 ds \Theta(s) [ 1 - 2 { \cal M}( { \bf r},s) ] q( { \bf r},s) q^{ \dagger}( { \bf r},s) \nonumber \\
\phi( { \bf r})&=& \frac{1}{ {2 \cal Q}} \int_0^1 ds q( { \bf r},s) q^{ \dagger}( { \bf r},s),  
\label{SPE1}
\eea
where the single-chain occupation variable, having performed 
the average over disorder, is simply given by $\Theta(s)=+1$ for $s<f$ and 
$\Theta(s)=-1$ for $s \ge f$ and where the single-chain polymer 
partition function is defined as
${ \cal Q}=\int d { \bf r} q( { \bf r},s) q^{ \dagger}( { \bf r},s)$. 
The two conjugated density fields in equation (\ref{SPE1}) are 
related to the forward propagator $q({ \bf r},s)$ via the modified 
diffusion equation,
\beq  
\frac{ \partial }{ \partial s}q( { \bf r},s)= \Big [ \frac{3}{ 2 a^2N} \nabla^2_{ \bf r} - { \cal W} ({ \bf r},s) \Big ] q( { \bf r},s) 
\label{SPE2}
\eeq
and a similar modified diffusion equation applies for the backward propagator 
$q^{ \dagger} ( { \bf r},s)$ . The self-consistent mean-field, given by 
$ { \cal W}({ \bf r},s)= \frac{1}{2} w( { \bf r})+ 
\frac{1}{2} v( { \bf r}) [ 1 - 2 { \cal D}({ \bf r},s) ]$, has two 
contributions, associated to the two conjugated density fields 
$w({ \bf r})$ and $v({ \bf r})$ above and depends on the disorder strength 
$p_{A,B}$ via
\bea
{ \cal D}( { \bf r},s)&=&\ln \big [ 1 + p(s) \big (e^{{ \cal G}({ \bf r},s)}-1 \big ) \big ]/{ \cal G}({ \bf r},s)  \nonumber \\
{ \cal M}( { \bf r},s) &=& p(s)e^{ { \cal G}( { \bf r},s)}/ \big [ 1+p(s) \big (e^{ { \cal G}( { \bf r},s)}-1 \big ) \big ]. 
\label{SPE3}
\eea
The two functions in equation (\ref{SPE3}) are obtained averaging with 
respect to the binomial form of the disorder each term in the partition 
function and resumming the associated series and depend on the average, 
auxiliary mean-field ${ \cal G}( {\bf r},s) $, containing four-propagator 
contributions \cite{Shakhnovich}, defined by
\beq
{ \cal G}( {\bf r},s) =\Theta(s) \big [ v({ \bf r}) - \frac{1}{ { \cal Q}} \int d { \bf r}'  v( { \bf r}') q( { \bf r}',s) q^{ \dagger}( { \bf r}',s) \big ]. 
\label{SPE4}
\eeq 
\begin{figure}[ht]
\begin{center}
\includegraphics[width=0.4 \textwidth,angle=270]{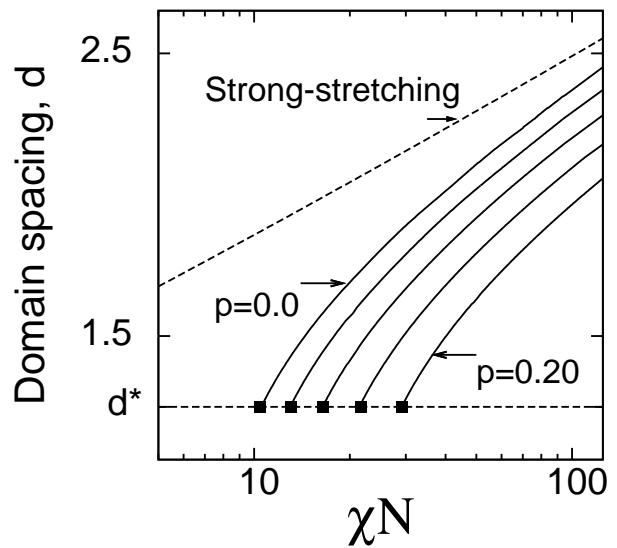}
\end{center}
\caption{Domain spacing $d$ as a function of the rescaled Flory-Huggins 
parameter $ \chi N$, for different values of the disorder strength, 
$p=0,0.05,0.1,0.15,0.20$. The dashed line indicates the asymptotic 
scaling form predicted by strong-segregation theory, 
$d \propto ( \chi N)^{1/6}$ \cite{Semenov}. At the order-disorder 
transition all curves, corresponding to different values of disorder 
strength, terminate at the same value $d^* \simeq 1.32$, as expected along 
the second order phase boundary separating the lamellar (L) and 
disordered phase (D) and indicated by the filled, black squares.}
\label{fig:fig3}
\end{figure}
In the absence of disorder, the saddle-point equations 
(\ref{SPE1})-(\ref{SPE4}) 
reduce to canonical self-consistent field theory \cite{Schick}. 
The modified diffusion equation (\ref{SPE2}) has been derived, within a replica 
symmetric ansatz for the many-body propagator 
$ q({ \bf r}_1, \cdots, { \bf r}_n,s)=\prod_{\alpha=1}^n q( { \bf r}_{ \alpha},s)$, following the principles of time-dependent mean-field 
Hartree-approximation \cite{Mezard}. 

We consider a system of block-copolymers with equal $A$-$B$-volume 
fraction and the particular case of a symmetric disorder 
strength within the two blocks, $p_{A,B}=p$. 
To solve numerically the associated saddle-point equations 
(\ref{SPE1})-(\ref{SPE4}) we used the real-space, one-dimensional 
Crank-Nicholson algorithm, in combination with a simple mixing technique. 
For any given value of the rescaled Flory-Huggins parameter we find 
the optimal value of the domain spacing $d$ that minimizes, according to a 
standard  one-dimensional minimization routine, the free-energy for different 
given values of disorder strength and rescaled Flory-Huggins parameter 
$ \chi N$. Typically we used $10^3$ points along both the spacial and temporal 
grids, though higher values $5 \times 10^3$ have been used to check 
accuracy of our results in the strong-segregation regime. 

In Fig. \ref{fig:fig1} we present the phase diagram of disordered 
block-copolymers with $A$-block volume fraction $f=1/2$. The solid thick 
line represents the second order phase boundary between the lamellar (L) 
and disordered (D) phases. As expected, lamellar order (L) is inhibited 
by the presence of disorder, but persists in the low temperature region, 
at large values of the disorder strength.     

In Fig. \ref{fig:fig2} we show the values of the order parameter $M$, for 
different values of disorder strength (top panel) and rescaled 
Flory-Huggins parameter $ \chi N$ (bottom panel), corresponding 
to horizontal and vertical scans in the phase diagram of Fig. \ref{fig:fig1}.
The order parameter $M$ is observed to depend on disorder. 
The location of the second-order phase boundary between the lamellar (L) 
and disordered (D) phases in the phase diagram Fig. \ref{fig:fig1} 
corresponds to the onset of points where the order parameter vanishes,  
as in Fig. \ref{fig:fig2}. 
The expected asymptotic form $M=(1-2p)$, shown as a dashed 
line in the top panel, is found to agree very well with the numerical 
results obtained within self-consistent field theory, in the intermediate 
segregation region $\chi N \gtrsim 60$ and at small values of disorder strength 
$p \lesssim 0.2$, while deviations form the strong-stretching results are 
described by self-consistent field theory, as in Fig. \ref{fig:fig2}. 
The asymptotic form we derived above for the order parameter 
$M$, scales as the effective block-size in the presence of uncorrelated 
disorder and numerical self-consistent field theory is observed to approach 
exactly this simple form.

In Fig. \ref{fig:fig3} we plot the domain spacing $d$, characterizing 
lamellar order (L), as a function of the rescaled 
Flory-Huggins parameter $ \chi N$. Domain spacing is observed to depend 
on the disorder strength, but not at the order-disorder transition, 
i.e. along the second-order phase boundary in the phase diagram, 
and as shown by the black filled squared boxes. 
In the strong-segregation limit the domain spacing of pure, 
non-disordered block-copolymers approaches the asymptotic form 
$d \propto ( \chi N )^{1/6}$ \cite{Semenov}. In fact, we observe that the 
domain spacing of disordered block-copolymer systems appears to 
cross-over to the same universal asymptotic form, shown by the dashed line.
The calculation of a global phase diagram, in the space of disorder 
strength $p$, block $A$ volume fraction $f$ and rescaled Flory-Huggins 
parameter $ \chi N$ can be addressed within the present theory, and the issue 
of the stability of other block-copolymer morphologies, beyond the 
one-dimensional lamellar (L) ordering here discussed, and the nature of the 
phase transition within the mean-field approximation and in the presence of 
quenched disorder is an open interesting problem. 

The important role of fluctuation effects, in the framework of one-loop Hartree 
approximation \cite{Stepanow}\cite{Brazovskii}, as well as momentum space 
renormalization-group theory \cite{Swift}, have been considered in the 
absence of disorder and corrections to the mean-field phase diagram of 
pure block-copolymer melts have been discussed \cite{Fredrickson2}
\cite{Zippelius}. The effect of thermal fluctuations in random copolymer 
systems have been discussed, within the weak-segregation approximation 
\cite{Shakhnovich2}. 
A model of block-copolymer systems with equal $A$-$B$-volume fraction $f=1/2$, 
in the presence of gaussian distributed Flory-Huggins interaction 
values, characterized by an average $\bar{ \chi}$ and disorder 
strength $\Delta \chi$, and the corresponding phase diagram have been 
discussed recently within renormalization-group theory \cite{Schmalian}. 
It has been argued that the presence of disorder changes 
the nature of the transition at a critical value of disorder 
strength and lamellar order becomes unstable at values of 
$\Delta \chi / \chi_{s}  \simeq 0.1$, where $\chi_s \simeq 10.495$ 
is the order-disorder transition in the absence of sequence disorder.

We observed that, within the mean-field theory approximation, lamellar 
ordering persists to large values of disorder strength, for any value 
of the Flory-Huggins parameter our algorithm can reach, and we 
in fact expect lamellar ordering to persist at low enough temperatures, 
for any value of disorder strength.

This research was funded by EPSRC under grant number EP/F068425/1. 
I acknowledge discussion with Alexei Likhtman and Pawel Stasiak.

\end{document}